\documentclass[conference]{IEEEtran}
\IEEEoverridecommandlockouts
\usepackage{cite}
\usepackage{amsmath,amssymb,amsfonts}
\usepackage{algorithmic}
\usepackage{graphicx}
\usepackage{textcomp}
\usepackage[table]{xcolor}
\usepackage{xcolor}
\usepackage{algorithm2e}
\usepackage{balance}
\usepackage{url}
\usepackage{tcolorbox}
\usepackage{multirow}
\usepackage{makecell}

\newcommand{\etal}{\textit{et al.}}
\newcommand{\ie}{\textit{i.e.,}}
\newcommand{\lcom}{{\sc lcom}}
\newcommand{\yalcom}{{\sc yalcom}}

\newcommand{\acc}{\rightarrow}
\newcommand{\mi}{\Rightarrow}

\begin{document}

\title{Do We Need Improved Code Quality Metrics?}


\author{\IEEEauthorblockN{Tushar Sharma}
	\IEEEauthorblockA{Siemens Technology\\
	Charlotte, USA \\
	tusharsharma@ieee.org}
\and
\IEEEauthorblockN{Diomidis Spinellis}
	\IEEEauthorblockA{Athens University of Economics and Business\\
	Athens, Greece \\
	dds@aueb.gr}
}

\maketitle

\begin{abstract}
The software development community has been using code quality metrics for the last five decades.
Despite their wide adoption, code quality metrics have attracted a fair share of criticism.
In this paper, first, we carry out a qualitative exploration by surveying software developers to
gauge their opinions about current practices and potential gaps with the present set of metrics.
We identify deficiencies including lack of \textit{soundness}, 
\ie\ the ability of a metric to capture a notion accurately as promised by the metric, 
lack of support for assessing software architecture quality, 
and insufficient support for assessing software testing and infrastructure.
In the second part of the paper, we focus on one specific code quality metric---\lcom{}
as a case study to explore opportunities towards improved metrics.
We evaluate existing \lcom{} algorithms qualitatively and quantitatively
to observe how closely they represent the concept of cohesion.
In this pursuit, we first create eight diverse cases that any \lcom{} algorithm must cover
and obtain their cohesion levels by a set of experienced developers and consider them as a ground truth.
We show that the present set of \lcom{} algorithms do poorly \textit{w.r.t.} these cases.
To bridge the identified gap, we propose a new approach to compute \lcom{}
and evaluate the new approach with the ground truth.
We also show, using a quantitative analysis using more than $90$ thousand types belonging to
$261$ high-quality Java repositories,
the present set of methods paint a very inaccurate and misleading picture of class cohesion.
We conclude that the current code quality metrics in use suffer from various deficiencies,
presenting ample opportunities for the research community to address the gaps.
\end{abstract}


\section{Background}
A software metric quantifies a characteristic or an attribute of a software product,
process, or project \cite{Fenton1997, timoteo2008, Kan2002}
that could be used later for assessment or prediction \cite{Fenton1997}.
\emph{Software product} metrics measure aspects such as size, complexity, and performance.
\emph{Process metrics} capture
aspects of software development processes,
such as the cost and effort required during the testing phase.
Finally, \emph{project metrics} concern project-level management
and resource allocation,
such as the number of active developers
or the schedule overrun.
In this paper, we focus on code quality metrics, a sub-category of product metrics.
We take a critical look at the existing metrics and
explore whether the software development and research community
require an improved set of metrics.

Code quality metrics have a long history,
which can be divided into three eras of research and practice.
In the first era, researchers mainly concentrated on size and complexity metrics,
such as the lines of code ({\sc loc}), cyclomatic complexity \cite{McCabe1976}, and the
Halstead metrics \cite{Halstead1977}.
The second era brought a plethora of interest in object-oriented metrics.
The most notable contributions of this period are the
Chidamber and Kemerer (C\&K) metrics suite \cite{Chidamber1994},
the {\sc mood} metrics \cite{Abreu1994}, and the design metrics by Martin \cite{Martin1995}.
We consider the year 2000 as the beginning of the third era of code quality metrics.
Research then changed focus
from proposing new metrics to utilizing code quality metrics
in applications regarding bug prediction \cite{Ambros2010},
maintainability prediction \cite{Li1993},
and code smell detection \cite{Lanza2007}.
Also, the era witnessed a wider adoption of metrics among practitioners
to track their code quality.

Commonly used code quality metrics could be classified in the following 
categories---\textit{size}, \textit{complexity}, \textit{coupling}, \textit{cohesion}, 
and \textit{inheritance} metrics.
Size metrics compute the size of program entities or constructs.
Particularly,
{\sc loc} was one of the first metrics that the software developers started using
in the late 1960s.

Halstead \cite{Halstead1977} proposed a metrics suite
to measure program size (termed as volume and vocabulary), based on rather primitive operator
and operand counts.
Function points \cite{Symons1988} compute size and complexity of a program
considering number of user inputs, outputs, inquiries, and external interactions.
Other commonly used size metrics are number of methods ({\sc nom}) in a class,
number of parameters ({\sc nop)} in a method, 
and number of classes ({\sc nc}) in an assembly, package, or a software system.

Cyclomatic complexity \cite{McCabe1976} is the most widely used complexity metric.
McCabe computed the complexity using $v(G) = e - n + 2$ where $e$ and $n$ refer to
number of edges and nodes in a control flow graph.
C\&K metric suite\cite{Chidamber1994} introduced weighted methods per class ({\sc wmc}) metric,
which calculates the complexity of a class by summing up the cyclomatic complexities of each method of the class.
Chidamber \etal{} \cite{Chidamber1994} also proposed coupling between objects ({\sc cbo}) metric,
which indicates the number of classes coupled with a given class.
The notion was further expanded to fan-in and fan-out metrics proposed by Henry and Kafura \cite{Henry1981}.
Fan-in and fan-out represent incoming and outgoing dependencies of a class respectively.
Similarly, Martin \cite{Martin1995} proposed afferent and efferent coupling metrics
to quantify incoming and outgoing dependencies for a module.
The Lack of Cohesion in Methods (\lcom{}) metric,
proposed in C\&K metrics suite,
captures the cohesion characteristic among a class's methods
based on their access to data members.
The category of inheritance metrics include depth of inheritance ({\sc dit}) and
number of children ({\sc noc}) \cite{Chidamber1994}.
Coleman \etal{} \cite{Coleman1994} proposed Maintainability Index ({\sc mi}) by combining
Halstead volume, cyclomatic complexity, lines of code, and percentage of comments.

\section{Deficiencies in Code Quality Metrics}
Software metrics have always been on a roller-coaster ride.
On one hand,
researchers and practitioners have adopted metrics not only to reveal quality characteristics
of their programs, but also to combine
existing metrics into new ones, and use these to study more complex phenomena.
On the other hand, metrics have drawn wide criticism.
The majority of this criticism is related to \textit{completeness}
and \textit{soundness}.

The first case covers the extent of
completeness of the implementation details provided to
compute the metrics independent from the programming language.
For instance,
implementation details of metrics in C\&K suite\cite{Chidamber1994} are missing,
leaving it up to one's interpretation \cite{timoteo2008, Mayer1999}.
Two examples of such deficiencies involve the lack
of concrete details for the implementation of the \textit{lack of cohesion in methods}
(\lcom{}) metric \cite{Mayer1999},
and the incomplete definition
of the \textit{coupling between objects} ({\sc cbo}) metric.
Particularly,
the definition of the {\sc cbo} metric does not clarify
whether both incoming and outgoing dependencies or only outgoing dependencies
should be used in the calculation.

\textit{Soundness} of a metric refers to the ability of the metric to accurately capture
the notion underlying its theoretical basis.
Shepperd and Ince \cite{Shepperd1994} presented their critical view on the Halstead,
cyclomatic complexity, and information flow metrics.
They wrote that the computing rules followed in the Halstead metrics are arbitrary and make use of
magic numbers without providing sufficient theoretical foundation.
For instance, programming time $T$ is computed through the formula $T = E/(f \times S)$,
where $E$ is referred to as programming effort, $f$ as seconds-to-minute ratio = 60,
and $S$ as  Stround number $= 18$.
Furthermore, these metrics use tokens and number of operators and operands,
which are considered too simplistic and
primitive, because they fail to capture control, data, and module structure \cite{Shepperd1994}.
Finally, Shepperd and Ince pointed out that
Halstead metrics were developed in the era of batch processing where software
systems were of a considerably smaller scale than today---often amounting
to a few hundred lines of code.
Similarly, other metrics such as cyclomatic complexity \cite{Shepperd1994, Mayer1999, Norman1999}
and maintainability index \cite{Sjoberg2012, avandeursen2014, Kuipers2007} are also
criticized for poor theoretical basis and validation.

A sound metric computing mechanism not only depends on the implementation,
but also on the crisp definition of the metric \cite{Fenton1997}.
From the implementation perspective,
different metric tools often produce different results for the same source code \cite{Sharma2018}.
More importantly, a metric is deprived from its usefulness if it does not measure what it intends to,
especially in commonly occurring cases.
SonarQube,\footnote{\url{https://www.sonarqube.org/}},
a widely used platform to measure and track code quality metrics and
technical debt,
identified that cyclomatic complexity fails to represent ``complexity'' of a snippet correctly
(for instance, in the presence of switch-case statements).
The platform has recently introduced a new metric,
\textit{i.e., cognitive complexity},
as an alternative to the
cyclomatic complexity metric \cite{SonarQubeCC}.

Similarly, despite the availability of many variants (\lcom{1} \cite{Chidamber1991},
\lcom{2} \cite{Chidamber1994}, \lcom{3} \cite{Li1993}, \lcom{4} \cite{Hitz1996},
\lcom{5} \cite{HendersonSellers1996}),
the lack of cohesion in methods (\lcom{}) metric is another example
where the intended characteristic, \textit{i.e.,} cohesion \cite{Etzkorn2004},
is not always correctly captured.
In RQ2 of this study (Section \ref{Section:RQ2}),
we compare the five variants of the \lcom{} metric
using different cases, and reveal their deficiencies in detail.

Apart from the issues of the existing metrics reported above,
the following concerns are also identified
regarding the commonly used code quality metrics.
First, the present set of code quality metrics is mainly designed for object-oriented programs.
Nowadays the software development community extensively uses programming languages,
such as Python and JavaScript,
that are not strictly object-oriented.
Hence, the present set of metrics is not applicable in its original form on such languages.
Second, the focus of code quality metrics has been limited to methods and classes.
There is hardly any metric at the level of architecture granularity.
For instance, akin to a complex method (inferred by using cyclomatic complexity metric),
it is difficult to comment anything about the complexity of a component or the entire software system.
Finally, recently,
sub-domains of software, including tests and configuration code
have become inseparable parts of the production code. 
The present set of metrics hardly support these sub-domains. 
Metrics such as code coverage for configuration and database code,
as well as coupling between the test or infrastructure code and the production code
could provide useful insight for the development community.

\section{Goal and Methodology}
The goal of this study is to first, understand the current practices in the
software development community
related to code quality metrics and
identify potential gaps.
Also, we aim to take one case study of a commonly used metric and explore
the soundness of the metric.

\begin{figure}[h]
	\centering
	\includegraphics[width=\linewidth]{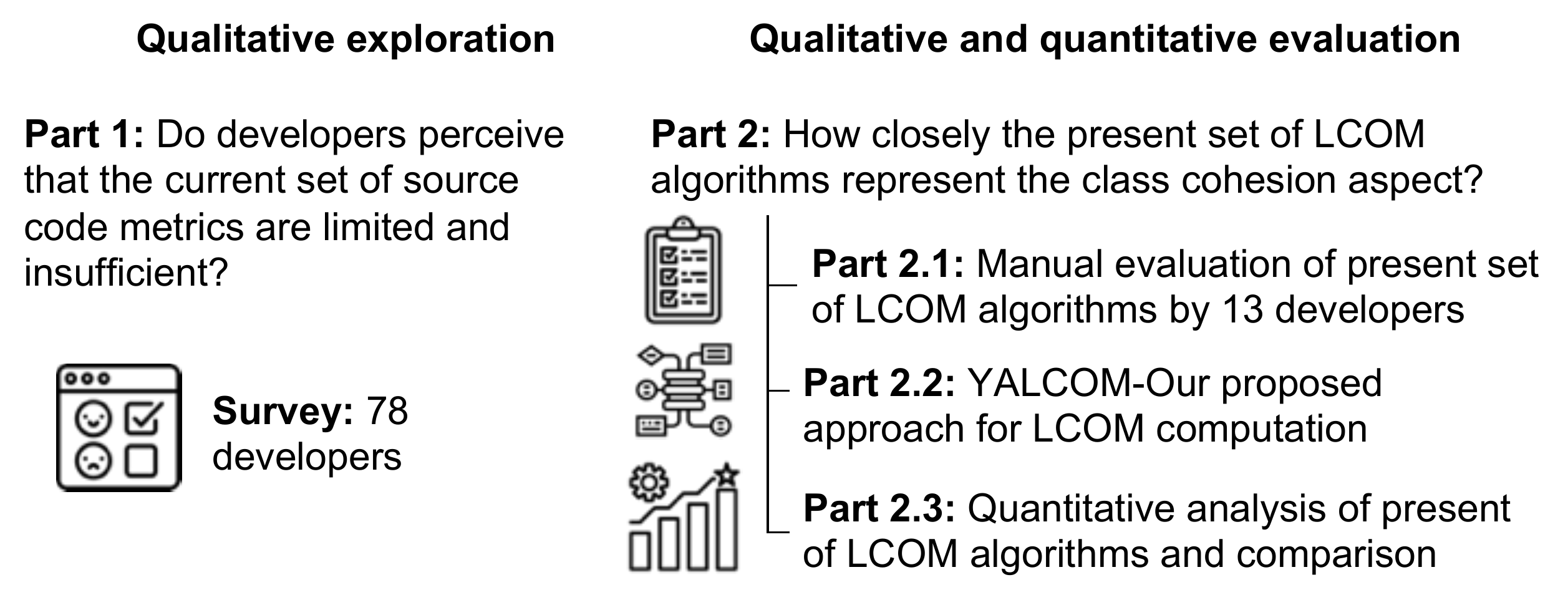}
	\caption{Overview of the method}
	\label{fig:overview}
\end{figure}

Towards the goals, we divide the study,
as outlined in Figure~\ref{fig:overview},
broadly into two parts.
In the first part, we carry out a qualitative exploration by surveying software developers.
The aim of the exploration is to gauge developers' opinions about current practices and potential
gaps concerning the present set of metrics.
For the case study, we considered cyclomatic complexity, Halstead metrics, and \lcom{} metrics.
Halstead metrics are not widely used; hence we dropped them from our consideration.
Cyclomatic complexity has been studied and criticized by many authors; hence, the deficiencies
of the metrics are well-known.
The \lcom{} metric has been also studied widely; 
however, their deficiencies have not been explored and evaluated by the community.
Therefore, we selected the metric for our case study.
In the second part we evaluate,
qualitatively and quantitatively, existing \lcom{} algorithms to observe their soundness
\ie\ how closely they
represent the concept of cohesion.
In this pursuit, we first create eight diverse cases that any \lcom{}
 algorithm must be tested against.
We first present these cases to experienced developers to obtain and establish the expected
values of cohesion in each of these diverse cases.
We compare the exiting set of \lcom\ algorithms and identify that 
the present set of \lcom{} algorithms do poorly \textit{w.r.t.} these cases.
To bridge the identified gap, we propose our approach of computing \lcom{} that we refer to as
{\sc yalcom} (Yet Another Lack of Cohesion in Methods).
Finally, we download $522$ Java repositories, analyze them to compute the \lcom{} metric for all
the classes in the analyzed repositories with existing methods as well as {\sc yalcom}.
We compare the outcome of all the considered algorithms by computing \textit{Euclidean distance}
among them
and deduce our observations based on the results of the experiments.

\subsection{Research Questions}
We explore two research questions in this study.
\begin{itemize}
\item [\textbf{RQ1.}] \textit{To what extent developers perceive the current set of code quality metrics
comprehensive and sufficient? } 
\end{itemize}
The first goal of this study is to explore the perception of software developers about  code quality
metrics.
Specifically, we aim to gauge their opinions about sufficiency and completeness
of the present set of commonly used metrics.

\begin{itemize}
\item [\textbf{RQ2.}] \textit{To what extent the existing set of \lcom{} algorithms capture the class cohesion aspect? }
\end{itemize}

As a case study,
we focus on a specific commonly used metric---\lcom{} and explore
to what extent the existing \lcom\ algorithms
capture the class cohesion aspect in the produced metric values.
We also aim to compare the existing algorithms with our method that we present in this study.


\section{RQ1. To what extent developers perceive the current set of code quality metrics
	comprehensive and sufficient?}

We discuss the mechanism, experiment, and the obtained results corresponding to the first
research question addressed in this study.

\subsection{Approach}
We designed a questionnaire to be used in an online survey
to gather the opinions of software developers and researchers.
The goal of the survey was to understand the practices followed by software developers
and identify potential gaps related to code quality metrics.

The first question asked the participants their
software development experience in the number of years.
The second question (multiple-choice) attempted to gather the commonly used metrics by the participants.
The next set of questions asked the participants whether the current set of metrics provides
sufficient insights about specific software engineering concerns (\textit{i.e.,} design, architecture,
test, and infrastructure).
We offered five-point Likert scale-based options
(\ie\ strongly agree, agree, neither agree nor disagree, disagree, and strongly disagree)
and asked them to choose one of the options. 
We also asked additional code quality metrics that they would like to have (open text).
The next two questions were targeted towards the extent to which the promised aspect is
represented and implemented correctly by the metrics (five-point Likert scale).
Finally, we asked the participants about their suggestions and feedback (open text).

We initially ran an internal pilot study,
which helped us refine the survey questionnaire according to the received suggestions,
and then publicly distributed the final survey questionnaire
through various social media platforms.
We kept the survey anonymous.
Table \ref{table:questions} presents the final set of questions in the survey.

\begin{table*}[ht]
	\rowcolors{2}{gray!25}{white}
	\begin{tabular}{p{0.03\textwidth}  p{0.92\textwidth} }
		\rowcolor{gray!50} \textbf{Q\#} & \textbf{Questions} \\\hline
1. &
Please mention your software development experience in years.\\

2.	&
Which code quality metrics [1] do you use/monitor for your source code. Check all that apply.\\
&
Lines of Code (LOC),
Halstead volume,
Average metrics (such as average method size and average number of methods per class),
Cyclomatic complexity (CC),
Weighted methods per class (WMC),
Halstead difficulty,
Halstead programming effort,
Lack of cohesion in methods (LCOM),
Coupling between objects (CBO),
Fan-in/Fan-out,
Afferent and efferent coupling (Ca/Ce),
Depth of inheritance tree (DIT),
Number of children (NOC),
Response for a class,
Comment percent,
Others (specify),
I/We do not use metrics,\\
&
1.	Stephen H. Kan. 2002. Metrics and Models in Software Quality Engineering (2nd ed.). Addison-Wesley Longman Publishing Co., Inc., Boston, MA, USA. 
\\

3.	& Do existing metrics provide the insights you need regarding the software's design?
\\

4.	&Do existing metrics provide the insights you need regarding the software's architecture?
\\

5.	&Do existing metrics provide the insights you need regarding the testing sub-domain of software development?
\\

6.	&Do existing metrics provide the insights you need regarding the infrastructure (operations, site reliability engineering, production engineering) sub-domain of software development?
\\

7.	&Which additional code quality metrics would you like to have?
\\

8.	&A metric's accuracy represents the extent to which the metric captures and represents the promised aspect. How often do you observe that some of the metrics are inaccurate in certain cases? For instance, value of a metric is reported 0 while you expected 0.5 due to the wrong implementation in the tool that you used.
\\

9.	&A metric's accuracy represents the extent to which the metric captures and represents the promised aspect. How often do you observe that some code quality metrics are inaccurate in certain cases? For instance, value of a metric is reported 0 while you expected 0.5 due to the specification of the algorithm used to calculate the metric (rather than a software implementation fault). 
\\

10.	&Do you have any comments, suggestions, reservations, feelings, or objections regarding the code quality metrics used commonly? Feedback to improve the survey is also welcome. Kindly provide your email address if you would like to receive compiled results of the survey (optional).
\\\hline

	\end{tabular}
	\caption{Questions in the developers' survey}
	\label{table:questions}
\end{table*}

\subsection{Results}

We received $78$ complete responses.
The participants belonged to various experience groups---no experience ($5\%$),
1--2 years ($12\%$), 
3--5 years ($17\%$), 
6--10 years ($21\%$), 
10--20 years ($32\%$), and
more than 20 years of experience ($13\%$).

\begin{figure}[h]
	\centering
	\includegraphics[width=0.8\linewidth]{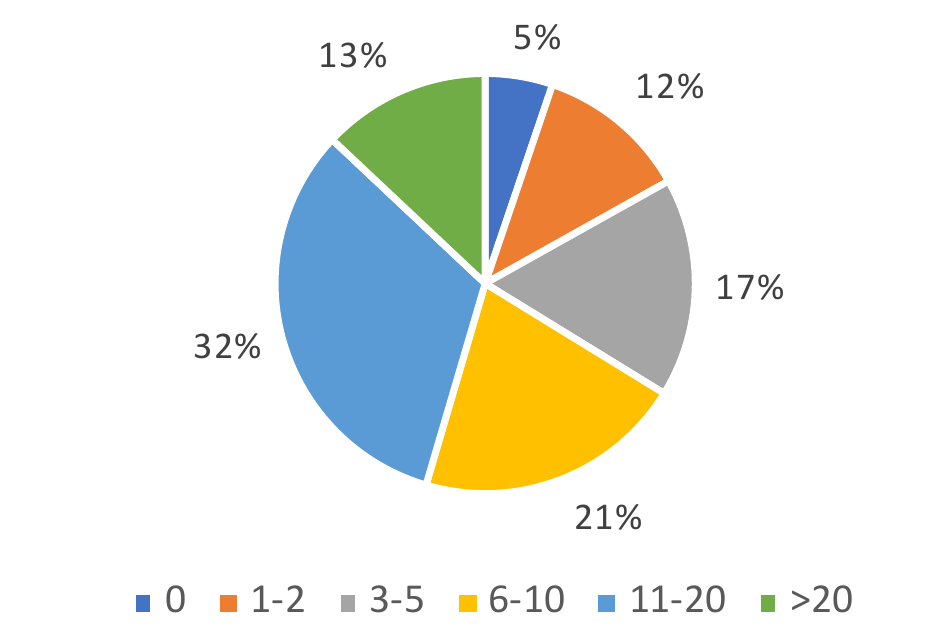}
	\caption{Software developer experience of the survey participants}
\end{figure}

In the second question,
respondents were asked to choose from a list of commonly used code quality metrics,
which ones they use and monitor on a regular basis.
As expected, lines of code ({\sc loc}) and cyclomatic complexity ({\sc cc}) are the most
commonly used metrics; $54\%$ and $52\%$ of the participants selected them, respectively.
Almost none of the participants use Halstead metrics.
Apart from the provided options,
participants further mentioned that they rely on test coverage and clone percentage.
Still, $26\%$ of the participants affirmed that they do not use any metrics.

\begin{figure}[h]
  \centering
  \includegraphics[width=0.9\linewidth]{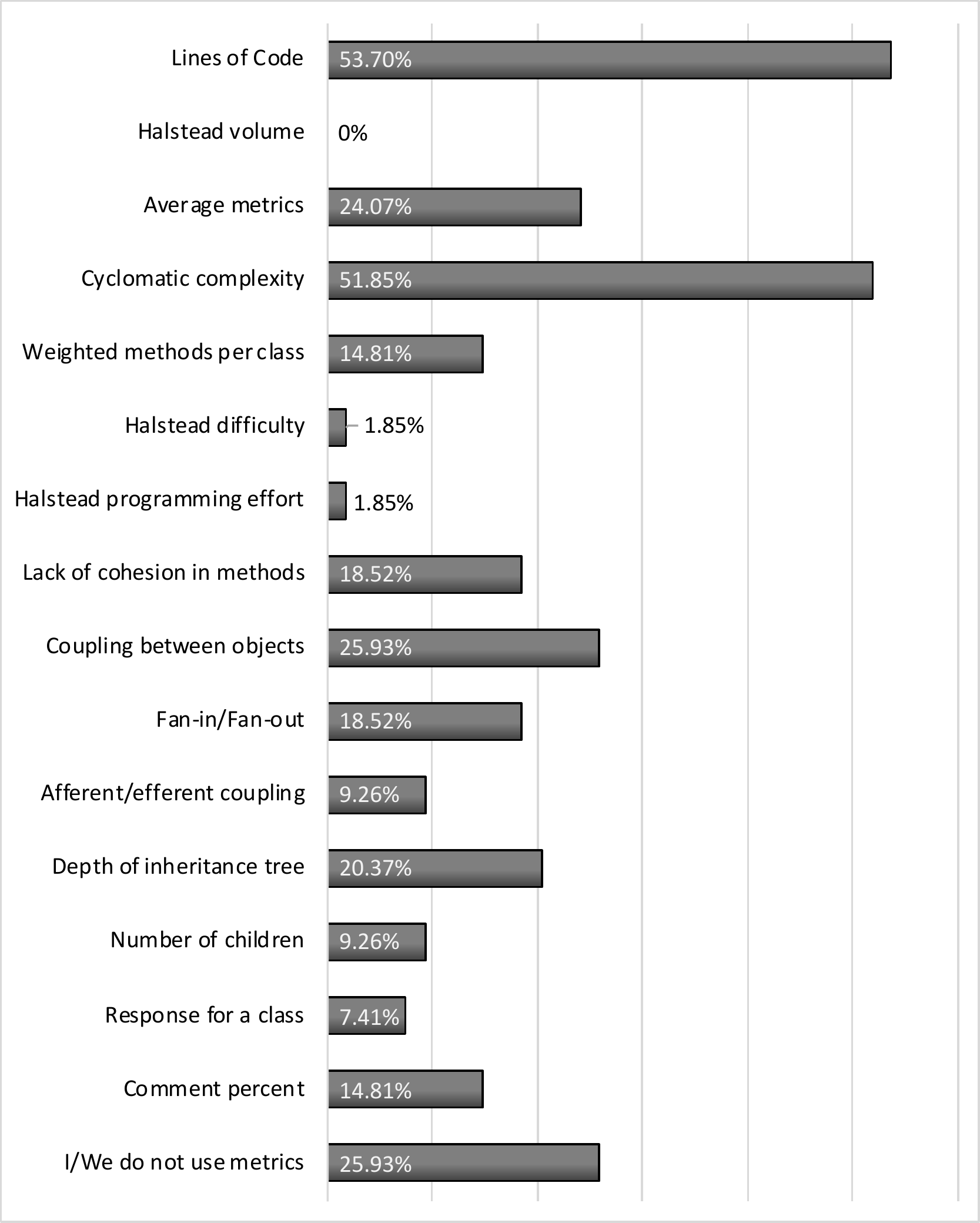}
  \caption{Commonly used metrics by the survey participants}
\end{figure}

Next, respondents were asked whether the current set of
metrics provides sufficient insight regarding software design.
The majority ($46\%$ agree and $6\%$ strongly agree) of the participants acknowledged
that they get insight about software design using the current set of metrics.
We asked a similar question about software architecture, and the majority 
($31\%$ disagree and $15\%$ strongly disagree) of the participants
stated that the current set of code quality metrics is inadequate
to measure quality of software architecture.
In the same context,
participants claimed that it is not easy to quantify architectural aspects.
As a respondent stated,
\textit{``you cannot easily distill architectural insight into numbers''}.
Other participants mentioned that the current set of metrics is only suited to measure
implementation and design aspects, but not architectural characteristics.

\begin{figure}[h]
	\centering
	\includegraphics[width=\linewidth]{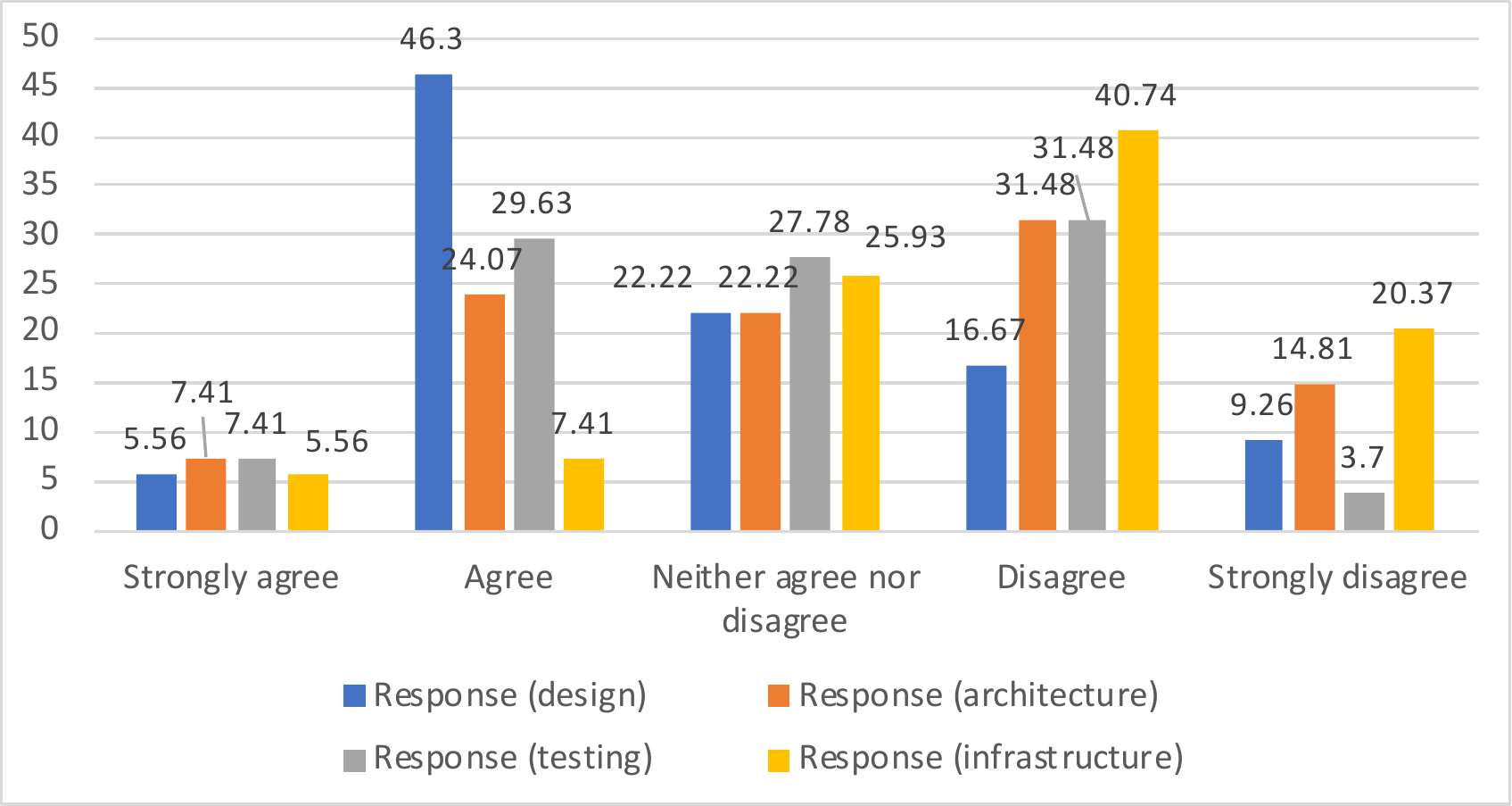}
	\caption{Survey responses toward assessing sufficiency of the commonly used metrics}
\end{figure}

The next question inquired
whether code quality metrics provide sufficient insight
for the sub-domains of software development testing and infrastructure.
Regarding the testing sub-domain, participants seemed divided.
The majority ($31\%$ disagree and $4\%$ strongly disagree) of the
participants stated that code quality metrics fall short with respect to testing.
In the words of a respondent, \textit{``test coverage is not enough''}.
On the other hand, another group of participants ($30\%$ agree and $7\%$ strongly agree)
claimed that the current set of testing-related metrics is sufficient.
The negative opinion was amplified for the infrastructure sub-domain.
A clear consensus emerged among $61\%$
($41\%$ disagree and $20\%$ strongly disagree) of the participants
who supported that code quality metrics are insufficient
to provide insight regarding infrastructure.
Undoubtedly, a lot of progress is expected toward measuring infrastructure code quality.

We got interesting and insightful answers in the response of the question that asked for additional
metrics that they would like to use.
The desired set of metrics mentioned, include architecture metrics
including specialized metrics for new architecture styles (e.g. microservices),
more context-sensitive metrics, `should I refactor this [method/class]' metric, code change frequency,
accidental complexity metrics, number of tests to cover all paths for different scopes,
the degree of adherence to a paradigm,
and metrics to show the extensibility of a module.
This set of desirable metrics coming directly from the practitioners may lead new efforts towards
novel code quality metrics. 

A metric's accuracy represents the extent to which the metric captures and 
represents the promised aspect.
Two additional questions were included
regarding metrics showing inaccurate values due to wrong
implementation and specification of the applied algorithm.
$32\%$ of the participants agreed that they often see inaccurate metric
values due to wrong implementation of the metric tools.
Another $39\%$ of the participants did not notice
any incorrect metric values.
Concerning algorithmic inaccuracies related to wrong specification,  
half of the participants expressed their ignorance 
(by choosing \textit{`I don't know'} option).

\begin{tcolorbox}
We summarize the results of the exploration below.
\begin{itemize}
\item 
The participants use a variety of code quality metrics, with
{\sc loc} and {\sc cc} being the most prevalent.
\item
The participants agreed that the present set of metrics provides insight for software design;
however, they recognized that the currently used metrics are insufficient to measure architecture quality
or to assess the testing and infrastructure sub-domains.
\item
The participants would like to see new metrics covering architectural aspects, code churn, and module extensibility.
\item 
Finally,
participants indicated that they often find inaccurate metric values due to wrong implementation.
\end{itemize}

\end{tcolorbox}


\section{RQ2. To what extent the existing set of \lcom{} algorithms capture the class cohesion aspect?}
\label{Section:RQ2}
This research question takes a concrete case of \lcom{} metric and attempts to gauge the soundness
of the existing set of methods to compute the metric.

\subsection{RQ2.1: Manual Evaluation of LCOM Algorithms} \label{RQ2.1}
\subsubsection{Approach}
In order to establish whether the commonly used set of \lcom{} metrics 
sufficiently capture the cohesion aspect of abstractions,
we handcrafted eight classes/interfaces representing different cases.
They are designed to cover various common cases involving interplay of
method calls, fields---their type (a class or an interface) and their accesses,
and inheritance that impact class cohesion
and may potentially reveal the deficiencies of the existing algorithms to compute \lcom{}.
Table~\ref{table:lcom} summarizes these cases;
their corresponding source code can be found in our online replication
package.\footnote{\url{https://anonymous.4open.science/r/5f0fd91a-8884-4977-8edb-47950f7a2f13/}}


\begin{table}[ht]
	\rowcolors{2}{gray!25}{white}
	\begin{tabular}{lp{0.4\columnwidth}p{0.35\columnwidth}}
		\rowcolor{gray!50}
		\textbf{Case} & \textbf{Class member relationships} & \textbf{Description} \\\hline
		Case1 & \makecell{$m_1\acc\{a_1, a_2, a_3\}$, \\ $m_2\acc\{a_2\}$, \\ $m_3\acc\{a_3\}$ }&  \\
		Case2 & \makecell{$m_1\acc\{a_1, a_2, a_3\}$,\\ $m_2\acc\{a_2\}$, \\ $m_3\acc\{a_3\}$} & $a_3$ is static \\
		Case3 & \makecell{$m_1\acc\{a_1, a_2\}$,\\ $m_2\acc\{a_1, a_2\}$,\\ $m_3\acc\{a_3\}$} &  \\
		Case4 & \makecell{$m_1\acc\{a_1, a_2, a_3\}$,\\ $m_2\acc\{a_1\}$,\\ $m_3\acc\{a_2, a_3\}$} & $a_1$ is in super class as a protected member \\
		Case5 & \makecell{$m_1\acc\{a_1\}$,\\ $m_2\acc\{a_2\}$,\\ $m_3\acc\{a_3\}$} &  \\
		Case6 & \makecell{$m_1\acc\{a_1, a_2, a_3\}$,\\ $m_2\acc\{a_2\}$,\\ $m_2\mi{}m_3$} &  \\
		Case7 & \makecell{$m_1, m_2, m_3$} & Type is an interface \\
		Case8 &  \makecell{$m_1, m_2, m_3, m_2\mi{}m_3$} & There are no fields (utility class) \\
		\hline
	\end{tabular}
	\caption{Cases covering various scenario concerning class cohesion;
		m$_i$ refers to a method, a$_j$ refers to an attribute (field) of a class,
		and relations `$\acc$' and `$\mi$' refer to \textit{access} and \textit{method invocation}, respectively.}
	\label{table:lcom}
\end{table}

To establish a ground truth, we put all of these cases
in the form of a survey and sent out the survey to experienced software developers in our network.
The survey had nine questions, one question per case showing the source code and one question
about their programming experience.
We gave four options to the participants for each case---the class/interface is cohesive,
the class/interface is partially cohesive,
the class/interface is incohesive,
cohesiveness could not be determined.

We consider five commonly used variants of the metric, \lcom{1}
\cite{Chidamber1991}, \lcom{2} \cite{Chidamber1994}, \lcom{3} \cite{Li1993},
\lcom{4} \cite{Hitz1996}, and \lcom{5} \cite{HendersonSellers1996}
for the experiment and implemented these algorithms.
We computed \lcom{} for all the eight cases using these  methods and compared them.

\subsubsection{Results}
\begin{figure}[h]
 	\centering
 	\includegraphics[width=0.55\linewidth]{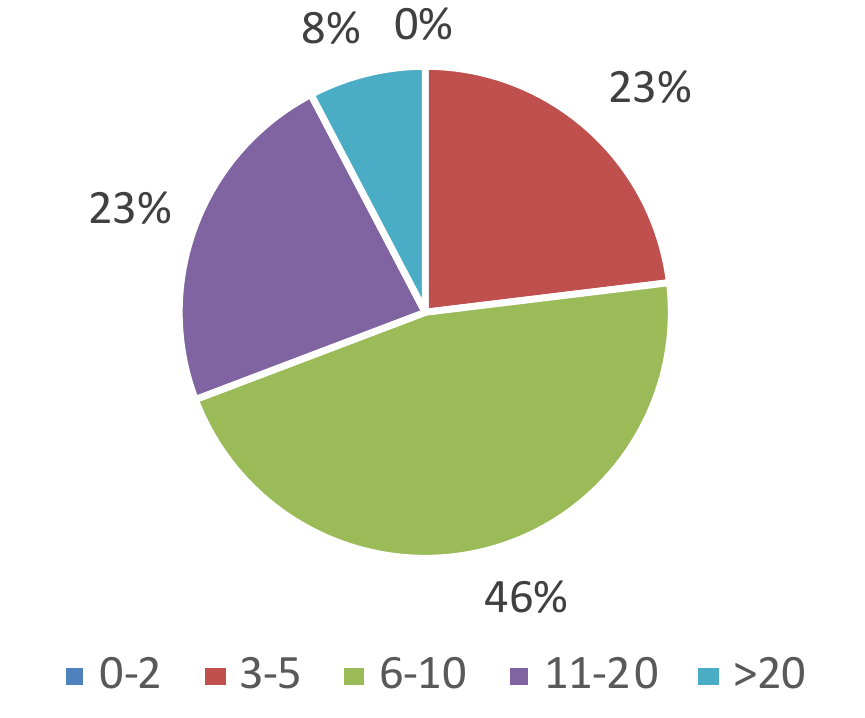}
 	\caption{Software development experience in years of the survey participants}
 	\label{fig:lcom_exp}
 \end{figure}

We received $13$ complete responses from our survey to establish ground truth.
Figure \ref{fig:lcom_exp} shows that the participants were highly experienced with software development.
They had $3$--$5$ years of experience at the minimum ($23\%$) while the majority of
the participants belonged to the $6$--$10$ years of experience category ($46\%$).

\begin{figure}[h]
	\centering
	\includegraphics[width=0.9\linewidth]{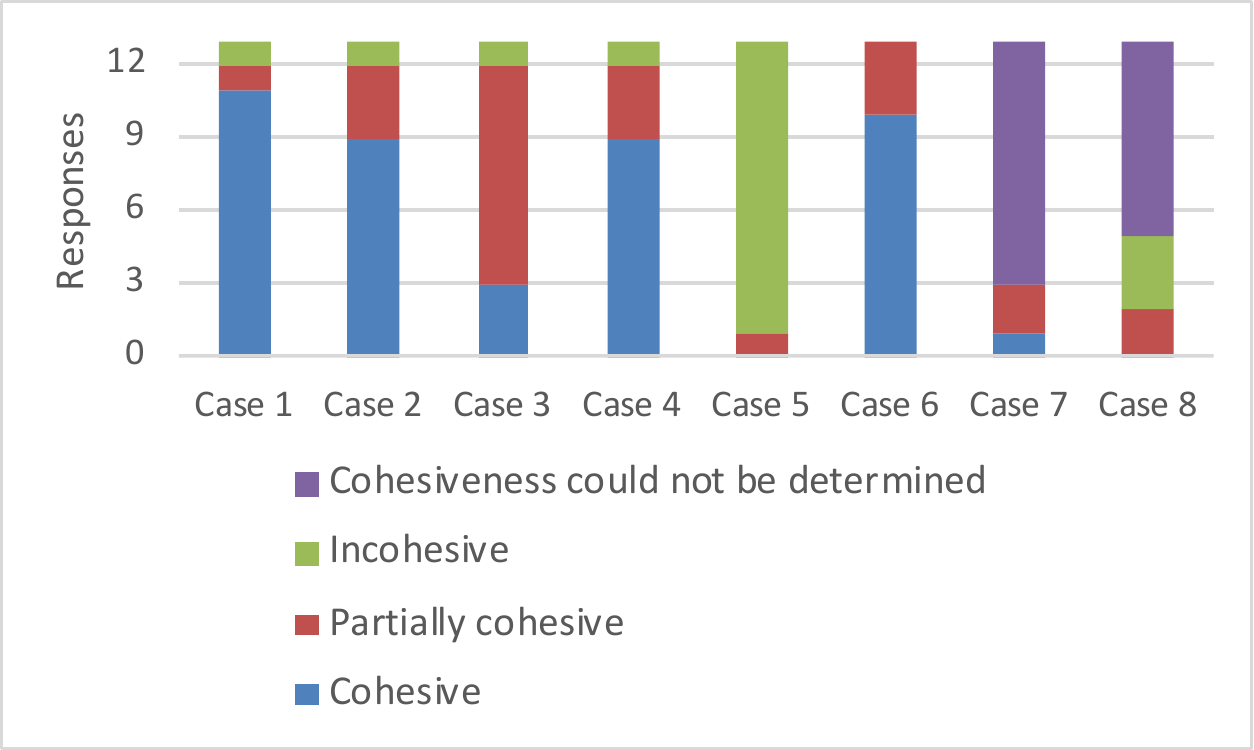}
	\caption{Responses to the survey for each case considered for establishing ground truth}
	\label{fig:ground_truth}
\end{figure}

Figure \ref{fig:ground_truth} summarizes the responses from the participants.
It is evident from the survey that \textit{Case1--2}, \textit{Case4}, and \textit{Case6} are considered
\textit{cohesive} by the majority of the participants.
Similarly, \textit{Case3} is tagged as \textit{partially cohesive} and \textit{Case5} is perceived
as \textit{incohesive}.
Participants marked \textit{Case7--8} as cases
where cohesiveness could not be determined because the available information is
not sufficient.
We consider the observations obtained from the survey as our ground truth.

Then, we  applied the existing \lcom{} algorithms and computed the metric value
for each of the eight cases.
Table \ref{table:lcom_compare} presents a comparison of the computed values
using existing methods with their corresponding expected ground truth.

\begin{table}[ht]
	\rowcolors{2}{gray!25}{white}
	\begin{tabular}{llllllll}
		\rowcolor{gray!50}
		\textbf{Case} & \textbf{L1} & \textbf{L2} & \textbf{L3} & \textbf{L4} & \textbf{L5} & \textbf{L} & \textbf{Ground truth}  \\\hline
		  1 & 1 & 0 & 1 & 1 & 0.67 & 0.00 & Cohesive  \\
		2 & 2 & 1 & 2 & 2 & 0.67 & 0.00 & Cohesive \\
		3 & 2 & 1 & 2 & 2 & 0.67 & 0.67 & Partially cohesive  \\
		4 & 2 & 1 & 2 & 2 & 0.50 & 0.00 &  Cohesive\\
		5 & 3 & 3 & 3 & 3 & 1.00 & 1.00 & Incohesive \\
		6 & 2 & 1 & 2 & 1 & 0.83 & 0.00 & Cohesive \\
		7 & 3 & 0 & 3 & 3 & 0.00 & -1.00 & Could not be determined \\
		8 & 3 & 0 & 3 & 2 & 0.00 & -1.00 & Could not be determined \\
		\hline
	\end{tabular}
	\caption{Comparison of the LCOM computed by existing methods with the ground truth.
		Here, L1--5 refer to the commonly used LCOM variants and
		L refers to the LCOM metric proposed in this study}
		
	\label{table:lcom_compare}
\end{table}

We derive
the following observations from the experiment.
\begin{itemize}
	\item \lcom{1--4} take into account only the instance attributes of a class,
	ignoring any static attributes.
	Although the dynamic properties of a static attribute differ
	from the instance attributes
	(for instance, static attributes can be accessed without creating an object),
	these attributes are part of the class they reside in.
	In the context of a metric that measures the similarity among class members,
	their dynamic property is irrelevant.
	Therefore, ignoring static attributes while assessing cohesion is inappropriate.
	\item 
	The existing \lcom{} algorithms fail to distinguish the cases
	where the metric cannot be measured, from the perfectly cohesive cases,
	by always emitting the lowest metric value in the former cases.
	For instance, \lcom{2} reports $0$ not only when the type under measurement is
	completely cohesive (Case1), but also when the type is an interface (Case7) and
	a utility class, \ie\ a class with no attributes (Case8).
	Such an approach produces the illusion to the user
		that all cases with a metric value of zero are cohesive,
	while in reality, the algorithm was not provided with enough information to measure the metric
	and the algorithm fails to communicate this to the user.
	\item
	Method invocations within a class show that methods are working together to
	achieve a goal and thus must be considered while computing \lcom{}.
	However, \lcom{1--3} and \lcom{5} do not consider method invocations
	to compute the metric.
	\item 
	Furthermore, the
	existing \lcom{} implementations focus on the common attribute access among methods
	within a class;
	however, they ignore common attribute access where the attribute is defined in a superclass.
	Classes are extensions of their superclasses, and it is very common to elevate data and method members
	to superclasses to avoid duplication among siblings.
	Hence, two methods that share attribute access or method invocation that is defined in a superclass contribute to
	cohesion and thus must be considered while computing the metric.
	\item
	Lastly,
	Fenton and Pfleeger \cite{Fenton1997} stated that a  metric may follow
	a suitable measurement scale (such as nominal, ratio, and absolute) depending on
	the aspect being measured.
	\lcom{1--4} measure cohesion on an absolute scale
	that may emit an arbitrary large number as the metric value
	making it almost impossible for the user to gain any insight from it.
	For instance, given that $m$ is the number of methods of a class,
	the maximum value that \lcom{2} may produce is $(m \times (m-1))/2$,
	which could be a considerable number for large classes.
	To facilitate metric interpretation and comparison,
	bounded concept such as cohesion must be better represented
	by a normalized value.
\end{itemize}

\vspace{2mm}
\begin{tcolorbox}
	We summarize the results of the experiment below.
	\begin{itemize}
		\item 
		Existing methods do not consider  relevant source code semantics
		such as static fields and method invocations to compute cohesiveness.
		Lack of such information leads to produce a metric value that is far from the real.
		\item
		The existing \lcom{} algorithms fail to distinguish the cases
		where the metric cannot be measured, from the perfectly cohesive cases.
		Such an approach produces an illusion to the user
		that all cases with a minimum metric value  are cohesive.
		\item
		The majority of the existing methods produce metric value on an absolute scale
		giving no clue to the user to interpret the value and giving no indication about what could be a good value to target for.
		\end{itemize}
	
\end{tcolorbox}

\subsection{RQ2.2: Our Proposed Approach for LCOM Computation}
\subsubsection{Approach}
To address the above-discussed deficiencies, we propose a new method 
referred to as \yalcom\ (Yet Another Lack of Cohesion in Methods)
to compute the \lcom{} metric.
Algorithm \ref{Algo:LCOM} presents the mechanism to compute the metric.

\begin{algorithm}
	\SetAlgoLined
	\KwIn{Type t}
	\KwOut{LCOM value}
	G = initialize a graph\\
	\eIf{isMetricComputable(t)}
	{
		G.addVertex(t.methods())\\
		G.addVertex(t.attributes())\\
		G.addVertex(t.supertype().attributes())\\
		\For{m : t.methods()}
		{
			G.addEdge(m, m.attributesAccessed())\\
			G.addEdge(m, m.methodInvocations())
		}
		d = G.disconnectedGraphs()\\
		\eIf{d $>$ 1}
		{
			LCOM = d/t.methods().size()
		}
		{LCOM = 0}
	}
	{LCOM = -1}
	
	\vspace{2mm}
	\SetKwFunction{isMetricComputable}{isMetricComputable}
	\SetKwProg{Fn}{Function}{}
	\Fn{\isMetricComputable{}}{\\
		\KwIn{t}
		\KwOut{True/False}
		\Begin{
			\If{t.methods().Count = 0 Or t.isInterface()}{return False}
			{return True}
		}
	}
	\caption{\yalcom{}---The proposed \lcom\ metric}
	\label{Algo:LCOM}
\end{algorithm}

\begin{figure*}[h]
	\centering
	\includegraphics[width=0.9\linewidth]{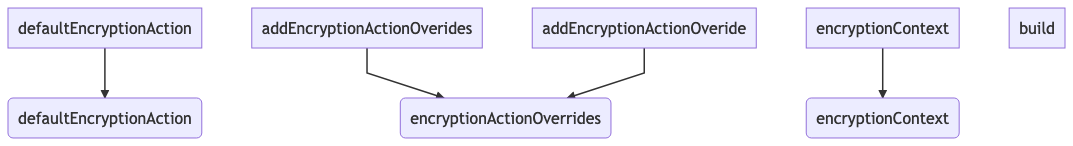}
	\caption{Class methods, fields, and relationship among them for class \texttt{Builder} belonging to AWS Dynamodb Encryption repository}
	\label{fig:example1}
\end{figure*}

\begin{figure*}[h]
	\centering
	\includegraphics[width=0.9\linewidth]{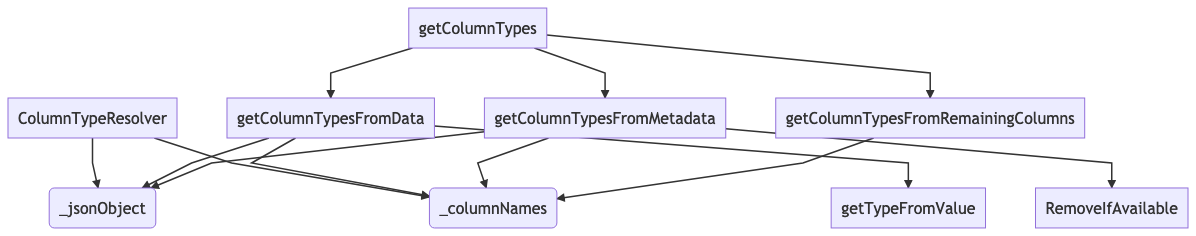}
	\caption{Class methods, fields, and relationship among them for class \texttt{ColumnTypeResolver} belonging to Apache Metamodel repository}
	\label{fig:example2}
\end{figure*}

The algorithm takes a type \ie\ a class or an interface as an input.
The algorithm returns $-1$ when 
the algorithm finds that the metric is not computable otherwise it returns a \lcom\ metric value $[0, 1]$.
The metric is not computable when the number of methods
is zero, or when the analyzed type is an interface.
The algorithm creates a graph where the methods and attributes of the class are treated as vertices.
Here, attributes from superclasses that are accessible from the class are also included.
Relationships, \ie\ field accesses and method invocations, among the methods and attributes, make the edges.
For example, if a method \texttt{m$_1$} accesses attributes \texttt{a$_1$} and \texttt{a$_2$} as well as 
calls method \texttt{m$_2$}, then the node corresponding to method \texttt{m$_1$} will have edges to
nodes representing attributes \texttt{a$_1$} and \texttt{a$_2$} as well as to method \texttt{m$_2$}.
Once the graph is constructed for the input class,
the algorithm finds the disconnected subgraphs of methods.
If the number of disconnected subgraphs is one,
then all the attributes and methods are connected to each other and hence the class is perfectly cohesive
(and thus assigned as $0$ as the metric value).
If the number of disconnected subgraphs is more than one,
it implies that there are many \textit{islands} of functionality within the class and hence
the class is not cohesive.
Here, the higher number of such subgraphs implies poorer cohesion.
We compute the metric by dividing the number of disconnected subgraphs by
the number of methods in the class.
Since the number of disconnected subgraphs cannot be more than the number of methods
(when none of the methods is associated with rest of the methods in the class),
the maximum value that the algorithm can produce is $1$.

\subsubsection{Results}
We evaluate the proposed algorithm on the established ground truth (refer to Section \ref{RQ2.1}).
Table~\ref{table:lcom_compare} shows the \lcom\ values (column L) produced by
the algorithm along with the expected values shown by the ground truth column for all cases.
The table clearly shows that the proposed algorithm computes the metric accurately for all the cases.

In addition to the ground truth cases, we manually checked the metric values computed by the
proposed method on a few Java repositories.
We present two examples one each from 
 AWS Dynamodb Encryption\footnote{\url{https://github.com/aws/aws-dynamodb-encryption-java}}
 and
Apache Metamodel\footnote{\url{https://github.com/apache/metamodel}}
repositories.
Figure \ref{fig:example1} presents the relationships of methods and attributes of class \texttt{Builder} in Dynamodb
Encryption repository.
Methods are shown by square-ended rectangles whereas attributes are shown in rounded rectangles.
Arrows show \textit{access} relationship from a method to an attribute and \textit{method invocation}
from a method to another method.
In this example, there are total of five methods and the method nodes
make four disconnected subgraphs and thus the computed \lcom\
would be $4/5=0.8$ which is considered an incohesive class.
Another example is from Apache Metamodel repository.
Figure \ref{fig:example2} shows methods and fields along with their relationships for class \texttt{ColumnTypeResolver}.
There are seven methods; they are well-connected with the rest of the nodes in the graph.
Hence the number of disconnected graphs is one and that implies that the class is perfectly cohesive with \lcom{}~$=0$.

\vspace{2mm}
\begin{tcolorbox}
	The ground truth test, as well as manual verification, provide sufficient indications that
the proposed \lcom\ algorithm produces metric values as expected by the ground truth and performs
superior compared to existing \lcom\
computation methods.
\end{tcolorbox}

\subsection{RQ2.3: Quantitative Analysis and Comparison}
This part of the exploration carries out a quantitative analysis to compare the existing algorithms
and their deviation from the values of the metric perceived by the developers.

\subsubsection{Approach}
To carry out a quantitative analysis, we downloaded high-quality Java repositories from GitHub,
compiled, and analyzed them to measure \lcom\ metric values for all the types in all the
analyzed repositories.

\vspace{3mm}
\noindent
\textbf{Download subject systems:} \\
We used the following protocol to identify our subject systems.

\begin{itemize}
	\item We use RepoReapers~\cite{Munaiah2017} to
	filter out low-quality and too small repositories among the abundant repositories
	present on Github.
	RepoReapers analyzed a huge number of GitHub repositories and evaluated
	each of the repositories on eight dimensions providing a fair idea about 
	their quality characteristics.
	These dimensions are architecture (as evidence of code organization),
	continuous integration and unit testing (as evidence of quality),
	community and documentation (as evidence of collaboration),
	history, issues (as evidence of sustained evolution),
	and license (as evidence of accountability).	
	\item We select all the repositories containing Java code where all
	the eight
	RepoReapers' dimensions have suitable scores.
	We consider a score suitable if it has a value greater than zero.
	
	\item Next, we remove the repositories that have less than $10$ stars as well as
	contain less than $1,000$ lines of code.
	
	\item Following these criteria, we selected and downloaded $522$ repositories.
\end{itemize}

A complete list of the selected Java repositories along with their
analyzed results can be found in our replication package.

\vspace{3mm}
\noindent
\textbf{Analyze subject systems:} \\
Our implementation of \lcom\ algorithms uses Eclipse JDT\footnote{\url{https://www.eclipse.org/jdt/}}
to prepare Abstract Syntax Tree and
resolve symbols.
It is necessary for the JDT libraries to have
compiled \texttt{.class} files along with source code files
to resolve the symbols and identify the various kinds of relationships correctly.
Therefore, we first compile the subject systems.
To carry out the compilation automatically,
we checked for the usage of one of the two commonly used build systems for Java
\ie{} \textit{Maven} and \textit{Gradle} and trigger the corresponding command
to compile the projects automatically.
This approach could compile a total of $261$ repositories;
we discarded rest of the repositories from further analysis.

We implement the existing methods to compute \lcom\ as well as our proposed algorithm.
We analyzed all the $261$ repositories and computed \lcom\ metric values for all the
types contained in the repositories.
We consolidated the obtained results, analyzed them, and documented our observations.

Euclidean distance is a common method to find the straight-line difference
in Euclidean space \cite{ONEILL2006} between two points.
Equation \ref{eq:euclidean_distance} shows the mechanism to compute collective Euclidean distance for
a series of points in Euclidean n-space \cite{ONEILL2006}.
The computed distance shows the extent to which two series are
similar or different.

\begin{equation}
d(\vec{u}, \vec{v}) = \sqrt{\sum_{n=1}^{n} (u_i-v_i)^2}
\label{eq:euclidean_distance}
\end{equation}

We compute the Euclidean distance between our proposed method and all the existing \lcom\ methods.
Since four out of five existing methods for \lcom\ produce metric values that are not
normalized, we computed their normalized values based on the min-max normalization
technique \cite{WITTEN2017}---we use the minimum and maximum values of \lcom\ values for each
specific repository individually for each considered method to produce their corresponding normalized values.
The normalization process brought all the metrics on the same scale to have an appropriate comparison.
We computed Euclidean distance for the normalized values also.

\subsubsection{Results}
We analyzed all the $90,029$ types belonging to $261$ repositories and computed \lcom\ metric value using
each of the existing \lcom\ algorithms along with our proposed method.
Then, we computed Euclidean distance for all the types between \lcom{1--5} and \yalcom{}.
Before computing the distance, we first identified and separated the types where the number of methods
is zero or the type is an interface.
We found that there are $15,356$ such types.
It implies that for approximately $17\%$ of the types,
the current \lcom\ methods were not able to compute the metric value
because either the type has no methods or
the type is an interface and hence there are no relationships
between methods and attributes.
Despite this, the existing methods show 
a perfect $0$ (or $1$ for \lcom{3--4}) indicating they are perfectly cohesive types
giving a false perception to the user of the metric about the class cohesion.

\begin{figure}[h]
	\centering
	\includegraphics[width=0.8\linewidth]{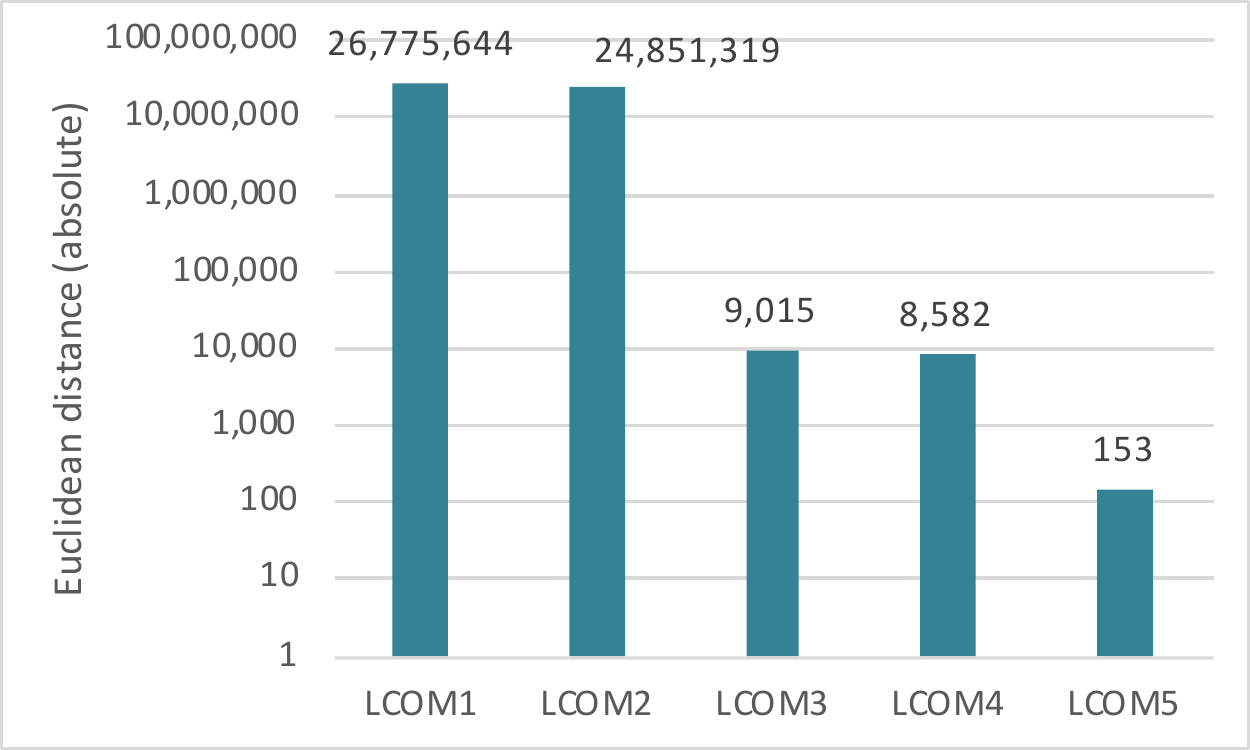}
	\caption{Euclidean distance (absolute) between  \lcom\ values generated by each of the LCOM1--5 and the proposed \yalcom{}}
	\label{fig:euclidean_abs}
\end{figure}

We computed the Euclidean distance for the rest of the types.
Figure \ref{fig:euclidean_abs} shows the computed absolute Euclidean distance between each of the
existing \lcom\ method and \yalcom{}.
It is evident from the figure that the metric values produced by the existing
methods are hugely different than the proposed method.
Relatively, \lcom{1--2} values are massively different from the proposed 
 approach because their \lcom\ values are on an absolute scale and hence, often,
 the produced values are arbitrary very large.
For example, the maximum metric values produced by \lcom{1} and \lcom{2}
are $22,221,184$ and $20,689,090$ for class \texttt{PlanProto}
belonging to Apache Tajo\footnote{\url{https://github.com/apache/tajo}}
repository.
We figured out that it is the case because the \texttt{PlanProto} class has
a total of $6,893$ methods belonging to its $207$ nested types where
the huge set of methods do not interact directly or indirectly with other methods
to form subgraphs that leads to the huge \lcom\ value. 
For the same class, \lcom{3} and \lcom{4} report $4,081$ and $4,076$ respectively
while \lcom{5} gives $0.99$.
Such arbitrary large metric values not only confuse the users since it is not known
whether a specific value is good or bad
but also fails to provide some actionable insight.
Table \ref{table:lcom_summary} summarizes some key characteristics of the analyzed \lcom\ values.

\begin{table}[ht]
	\begin{tabular}{lrrrr}
		\textbf{Algorithm} & \textbf{Maximum} & \textbf{Minimum} & \textbf{Median} & \textbf{Average}  \\\hline
		\lcom{1} & 22,221,184 	& 0	&	4	&	960.77 \\
		\lcom{2} & 20,689,090	& 0 	& 0	& 	817.23 \\
		\lcom{3} & 4,081	& 1 	&2	&	6.46	\\
		\lcom{4}&	4,076	&	1&	2	&	5.26\\
		\lcom{5}&	2	&	0&	0.5	&	0.44\\
		\yalcom{}&	1	&	0&	0.22	&	0.36\\
		\hline
	\end{tabular}
	\caption{Characteristics of the \lcom\ values generated by all considered \lcom\ methods derived from our quantitative analysis}
	
	\label{table:lcom_summary}
\end{table}

\begin{figure}[h]
	\centering
	\includegraphics[width=0.8\linewidth]{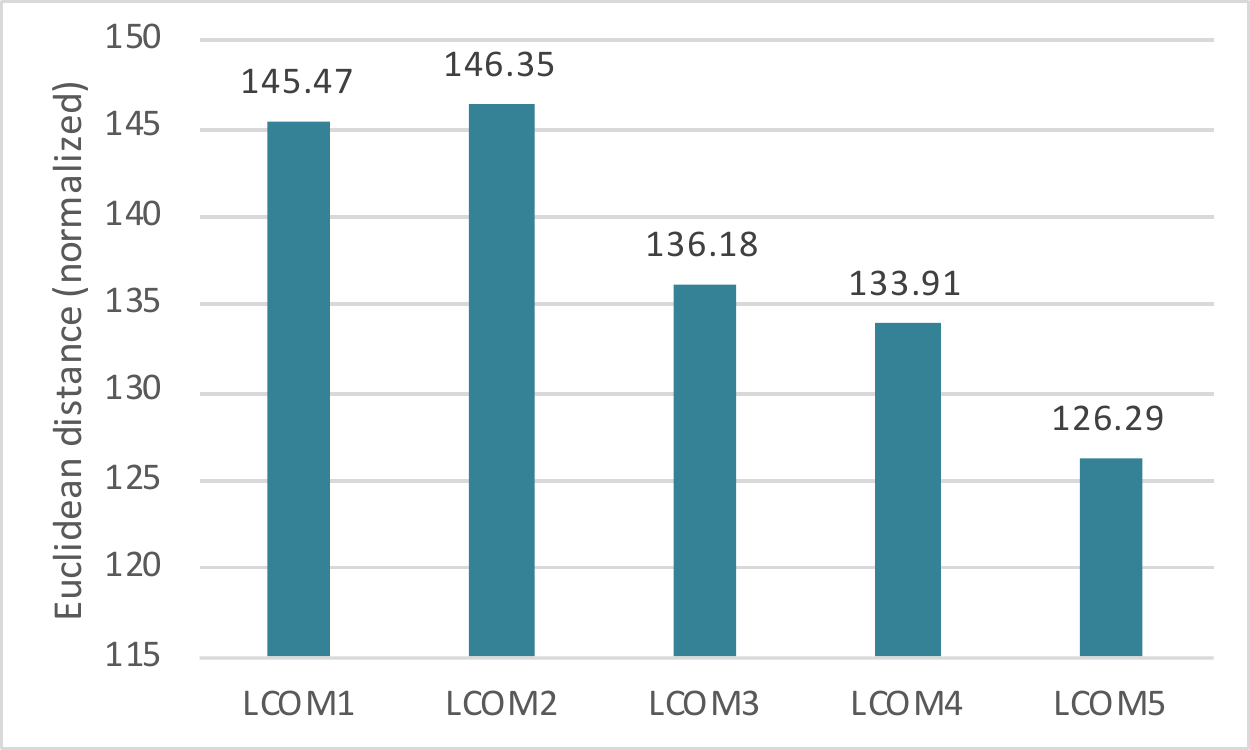}
	\caption{Euclidean distance (normalized) between \lcom\ values generated by each of the LCOM1--5 and the proposed \yalcom{}}
	\label{fig:euclidean_norm}
\end{figure}

One may argue that it is not fair to compare the absolute values of the \lcom{},
though generated correctly by definition, from existing methods where the values are not normalized.
To observe the difference and compare them on the same scale,
we normalized the metric values in the range of [0, 1].
Figure \ref{fig:euclidean_norm} presents the comparison of all the considered \lcom\ methods
after the normalization.
Though the scale is different, the figure reflects a similar pattern as we see
in  Figure~\ref{fig:euclidean_abs}.
\lcom{5} computes the closest values to \yalcom\ compared to the rest of the existing methods.
However, in summary, all the existing methods produce \lcom\ values very different
from the expected values.

\begin{tcolorbox}
We summarize the findings of the quantitative analysis as follows.
\begin{itemize}
	\item 
	The existing \lcom\ algorithms report a perfect score indicating the type is perfectly cohesive
	even when there are not enough details available to deduce the conclusion.
	We found $17\%$ of the types belonging to this category.
	The deficiency gives a false perception to the user about the cohesiveness of the class.
	\item 
	The \lcom\ values computed from existing methods produce values that are
	far from the values produced by the validated method.
	Hence, the existing methods do not capture the notion of the metric accurately. 
\end{itemize}
\end{tcolorbox}

\section{Discussion and Implications}

We identify the following implications for the software engineering community.
\begin{itemize}
	\item 
	For software engineering researchers, the exploration presented in the study
	reveals ample opportunities to fill the identified gaps by, for instance,
	proposing relevant and effective code quality metrics.
	Furthermore, the study paves the way to look for alternatives to deterministic and traditional
	 code quality metrics.
	Additionally, the presented study invites researchers to gauge the effectiveness and soundness
	of  existing metrics by presenting a case study of the \lcom\ metric.
	Similar studies can be explored for other commonly used metrics. 
	\item 
	Software tool developers and vendors may take a cue from the study and innovate
	novel mechanisms to perceive the specific quality aspects of a software system.
	Also, they could refine the existing metric implementations that better
	represent the measured aspect.
	\item 
	The take away for a developer from this study is that the metrics that are being used today
	do not necessarily represent the aspect that the metric claim to and hence
	the over-reliance on strict quality processes based only on such metrics may not be needed.
\end{itemize}

Naturally, the discussed deficiencies and criticism can be addressed by proposing new mechanisms and metrics.
Alternatively,
advances in machine learning technologies and their practical feasibility
have introduced an additional twist
regarding code quality metrics.
Traditionally,
code quality metrics based on intuition, theory, or empirical evidence,
offered a simple shortcut to assess source code quality.
Essentially, researchers find rules and heuristics, and justify them with their theory
and empirical evidence.
On the contrary, machine learning-based methods look at the data and derive
their heuristics from them.
This implies that the community may come up with advanced machine learning-based
quality estimation models,
that are not necessarily derived from existing metrics
but are trained on a gamut of software engineering development data.
Such an approach offers advantages, issues, and challenges.
On the one hand, machine learning approaches may offer increased accuracy
by flagging quality problems that are not discernible through simple metrics.
On the other hand, machine learning results might be less actionable,
since they may point to questionable code without providing an explanation
on why the specific code is problematic.
And yet, a positive aspect of the machine learning approach may be
that developers will be less likely to game the system
by manipulating the metrics and their thresholds,
rather than actually improving the software's quality.
A widely known challenge concerning machine learning-based methods is the
availability of well-curated labeled data, as
the success of these methods depends heavily on them.
Our take is that the software engineering community should
consider machine learning approaches regarding the judgment of
code quality as another type of metric,
one that, in common with existing metrics, should stand or fall based on
its usefulness backed by empirical evidence.

\section{Replication Package}
We provide the following in our anonymized online replication package\footnote{\url{https://anonymous.4open.science/r/5f0fd91a-8884-4977-8edb-47950f7a2f13/}}.
\begin{itemize}
	\item Source code of all the cases used to establish the ground truth
	\item Implementation of existing \lcom\ algorithms as well as the proposed algorithm
	\item A list of Java repositories used for quantitative analysis
	\item Results generated from the \lcom\ implementations for all the considered repositories
	\item Consolidated results containing \lcom\ absolute as well as normalized values for all the types
\end{itemize}

\section{Threats to Validity}
\textit{Construct validity} concerns the appropriateness of observations and inferences
made on the basis of measurements taken during the study.
We propose a new algorithm to compute the \lcom\ metric and derive observations
based on the algorithm and the corresponding implementation.
We validated the proposed algorithm against a ground truth that
itself is prepared from the inputs provided by the experienced developers.
In addition, we carried out random manual testing to ensure that the implementation
is working as intended.

We derive a set of cases to cover different scenarios of a class configuration with methods,
attributes, and their relationships.
Though we tried to cover all common cases but it is possible that we have missed other
relevant cases.
We encourage researchers to extend the presented cases so that a benchmark to measure
cohesion can be set up.

\textit{External validity} concerns the generalizability and repeatability
of the produced results.
We carried out our experiments on a large number of repositories to keep the results
generic.
We have made all the scripts as well as data produced in the experiments available
to promote reproducibility.

\textit{Internal validity} refers to the validity of the research findings.
It is primarily concerned with controlling the extraneous variables and external 
influences that may impact the outcome. 
Our online survey carried out by distributing it over the Internet via social media.
Given the anonymous nature, we do not know whether the participants were indeed software developers
and actually represent a common perception of a typical software developer.
However, given a relatively large number of participants ($78$), 
we rely on the overall trend that we observe from the obtained choices.

\section{Conclusions}
We provide an overview of the common code quality metrics in use since the 1960s,
present criticism derived from the literature, and outline
deficiencies in the currently used code quality metrics.
We also present a developers' perspective,
which clearly shows that the current set of
metrics is not sufficient for a variety of reasons.
The exploration reveals deficiencies in the current set of code quality metrics
such as poor support to assess software architecture quality as well as to assess
testing and infrastructure aspects of software systems.
As a case study, we provide a detailed qualitative and quantitative 
analysis of presently used \lcom\ metric algorithms
and point out specific deficiencies.
We proposed a new algorithm to compute \lcom\ and show that it produces metric values
as expected from the established ground truth.
We also show that the presently used \lcom\ algorithms are not sound.

In the future, we would like to propose traditional  code quality metrics for sub-domains,
such as infrastructure,
where sufficient metrics are not yet invented.
We are also interested to explore machine learning approaches to spot
source code entities where code quality is not optimal.


\balance

\bibliographystyle{IEEEtran}
\bibliography{metrics_arxiv}
\end{document}